\documentclass[11pt]{article}
\usepackage{blois,epsfig,graphicx,amsmath}

\usepackage[numbers]{natbib}
\usepackage{color}
\def\be{\begin{equation}}
\def\ee{\end{equation}}
\def\bea{\begin{eqnarray}}
\def\eea{\end{eqnarray}}

\begin{document}
\vspace*{4cm}
\title{The Advanced LIGO Gravitational Wave Detector}

\author{ S. J. Waldman,  for the LIGO Scientific Collaboration }

\address{LIGO Laboratory, Kavli Institute for Astrophysics and Space
  Research, MIT NW22-270,  185 Albany St., Cambridge MA 02139}

\maketitle
\abstracts{The Advanced LIGO gravitational wave detectors
  will be installed starting in 2011, with completion scheduled for
  2015.  The new detectors will improve the strain sensitivity of
  current instruments by a factor of ten, with a thousandfold increase
  in the observable volume of space.  Here we describe the design and
  limiting noise sources of these second generation, long-baseline,
  laser interferometers. }

\section{Introduction}
Einstein's general theory of relativity predicts the existence of
gravitational waves (GWs), oscillations in the space-time metric that
propagate at the speed of light. The Laser Interferometer
Gravitational-Wave Observatory, LIGO, is designed to detect and study
astrophysical GWs, with the promise of studying qualitatively new
physics and astrophysics.~\citep{Abramovici:1992p2546} In particular,
the direct detection of GWs will provide information about systems in
which strong-field gravitation dominates, a virtually untested regime
in which space-time curvature self-interacts.  Such GW sources include
compact binary coalescences in which a neutron star or black hole
binary system inspirals together, coalescing to form a black hole; the
stellar core collapse thought to power Type II supernovae; rapidly
rotating asymmetric neutron stars; and possibly cosmic-scale processes that
produce a stochastic background of GWs.~\citep{Cutler:2002p7832}

In the past few years, the first generation of long-baseline
gravitational wave detectors has successfully operated at or near
design sensitivity.  In collaboration with the Virgo 3~km and the GEO
600~m interferometers,~\citep{Acernese:2008p9535, Willke:2002p3770}
LIGO anchors a worldwide network of instruments in search of the first
direct detection of gravitational waves.  The LIGO detectors operated
from November 2005 to October 2007, with joint data taking with Virgo
starting in May 2007.  The data is currently being analyzed for GW
signals from inspiraling binary systems, burst sources, a stochastic
GW background, and rapidly rotating neutron stars.  The status of
these searches and their astrophysical importance is discussed by
other authors in these proceedings.  

This article focuses on the next generation of LIGO interferometers,
in particular Advanced LIGO, currently being designed and assembled at
two sites in the United States.  The Livingston Parish, Louisiana
observatory will operate a single interferometer, L1, with 4~km long
arms while the Hanford, Washington observatory will operate two 4~km
interferometers within a common vacuum envelope, H1 and H2.  The
second generation Advanced LIGO detectors will improve the sensitivity
of ground-based gravitational wave detectors by an order of magnitude
over current detectors.  A preliminary Advanced LIGO design was
described in Ref.~\citep{Fritschel:2003p9356}, here we provide an
overview of the final design as construction begins.  We first
describe the Advanced LIGO optical configuration, then follow with a
description of the dominant noise terms and anticipated sensitivities.
Finally, we conclude with comments on the initial tests of Advanced
LIGO and progress towards the first science runs.

\section{Advanced LIGO Optical Design}
\label{sec:advanced-ligo-design}

From a detector perspective, gravitational waves can be thought of as
a quadrupole strain of space, $h = \delta L / L$, which can be probed
by monitoring the relative positions of inertial test masses with
light.  Equivalently, in a fixed Lorentzian frame, gravitational waves
create a tidal force, a force proportional to the distance from a
chosen origin.  As with electromagnetic waves, GWs are transverse
waves that travel at the speed of light.  Unlike electromagnetism, GWs
are constrained by mass and momentum conservation to be quadrupolar:
the strain (or tidal forces) contracts along one transverse dimension
while expanding the orthogonal dimension.  Also unlike
electromagnetism, GWs are very weak and interact very weakly as they
propagate through space; detectable GWs are generated only by the
coherent acceleration of stellar masses at relativistic velocities.  The
strongest nearby sources will produce strains on Earth no larger than
$h \approx 10^{-21}$.  Finally, it's worth noting that GW detectors
measure the amplitude of a GW (as opposed to the power) so that the
observed volume of space scales cubicly with the detector sensitivity.

\begin{figure}[tb]
  \centering
  \includegraphics[height=4in]{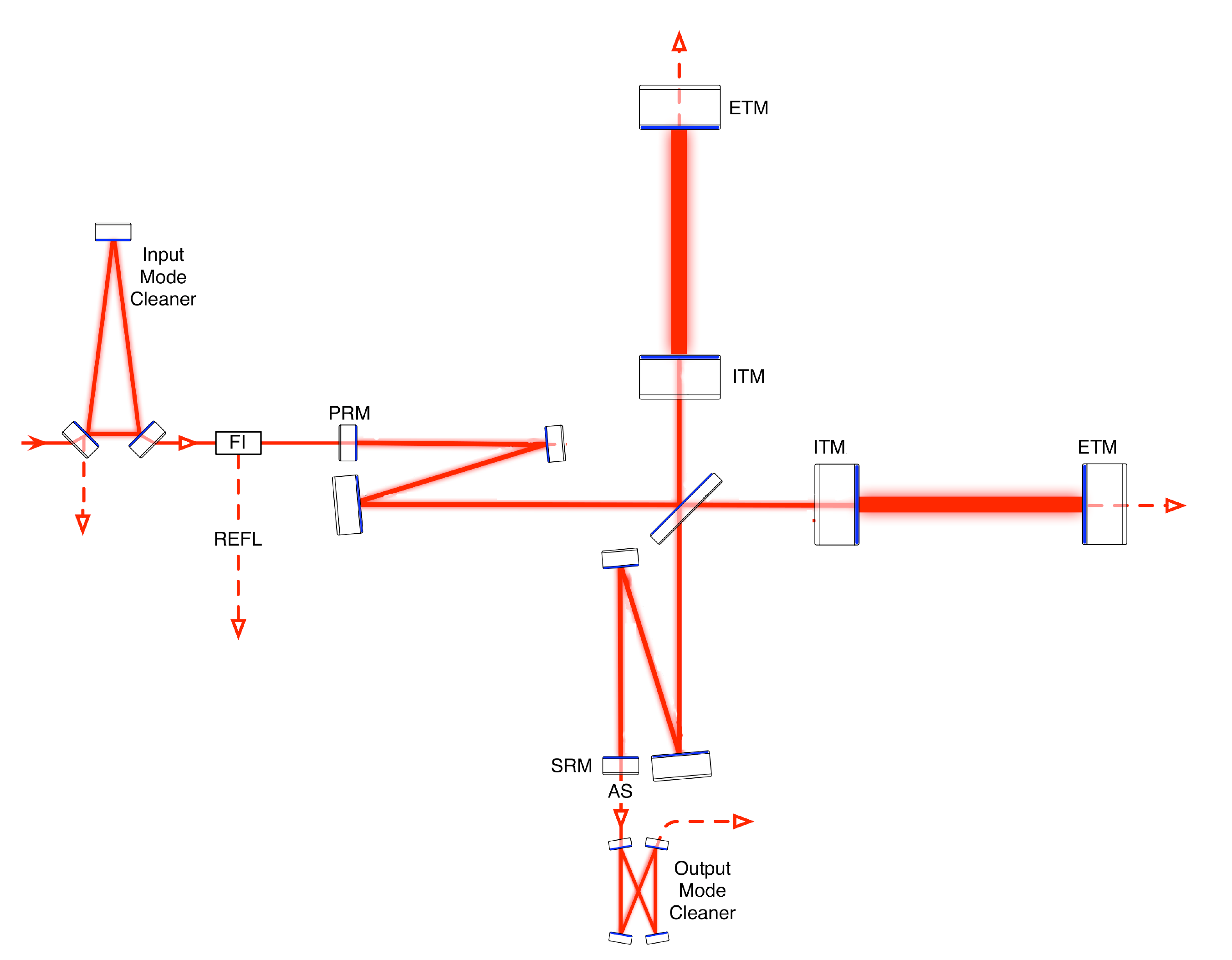}
  \caption{The Advanced LIGO optical layout. The triangular suspended
    input mode cleaner filters frequency and amplitude noise from the
    laser (not shown) and provides a stable input beam.  The Faraday
    Isolator (FI) isolates the laser from the interferometer reflected
    beam (REFL) used to control the laser frequency.  The 4~km long
    Fabry-Perot arm cavities are formed between the ITM and ETM test
    masses.  The Power Recycling Mirror (PRM) and Signal Recycling
    Mirrors (SRM) form folded cavities discussed in the text.  The GW
    signals are carried by the light transmitted through the Output
    Mode Cleaner at the Anti-Symmetric (AS) port.}
  \label{fig:layout}
\end{figure}

The Advanced LIGO optical design consists of a Michelson
interferometer with Fabry-Perot arm cavities, a power recycling cavity
and a signal recycling cavity as shown in Fig.~\ref{fig:layout}.  The
Michelson topology is well matched to the quadrupole strain: a
properly oriented, linearly polarized GW propagating normal to the
interferometer plane generates a positive strain along one arm, a
negative strain along the other and vice versa, oscillating in time.
In the Advanced LIGO configuration, the arm lengths are controlled so
that Michelson interferometer reflects the input laser beam back
towards the laser while the anti-symmetric port is dark. The
differential motions of the two arms -- the GW signal --
constructively interfere at the beam splitter's Anti-Symmetric port,
labeled ``AS'' in Fig.~\ref{fig:layout}.  The common mode signals
generated by common mode motion of the end mirrors, by laser frequency
noise, and by laser intensity noise constructively interfere at the
beam splitter's Symmetric port; to first order, this configuration
eliminates laser technical noise couplings to the GW signals.

Fabry-Perot cavities in the interferometer arms defined by the
partially reflecting Input Test Mass (ITM) and high reflectance End
Test Mass (ETM) resonate the input laser light to increase the power
in the arms. Similarly, the partially reflecting Power Recycling
Mirror (PRM) resonates the light that returns toward the laser from
the beam splitter's symmetric port.  Together the arm and power
recycling cavities build up the laser power in the arms by a factor of
$\simeq 6000$. The power recycled Fabry-Perot Michelson topology is
identical to Initial LIGO and has been described in detail in
Ref.~\citep{Abbott:2009p9436}.

Advanced LIGO has several significant changes in the optics relative
to Initial LIGO: the Signal Recycling Mirror (SRM) is added to the AS
port of the interferometer to form a signal recycling cavity, the
power and signal recycling cavities have a stable geometry, the GW
signal is detected using DC readout, and the laser power is increased.
The SRM forms a resonant cavity for the differential mode signal,
altering the interferometer
dynamics.~\citep{Meers:1988p8021,Mizuno:1993p10166} The impact on the
interferometer quantum noise is discussed further in the
\S\ref{sec:quantum} below.

In Initial LIGO, the $\approx 10$~m long power recycling cavity was
formed by mirrors having $\ge 10$~km radii of curvature, effectively a
flat-flat resonator geometry.  That configuration was extremely
sensitive to changes in the curvatures caused by unavoidable thermal
lensing.  To reduce the thermal sensitivity, the Advanced LIGO signal
and power recycling cavities are each formed by a folded chain of
three curved mirrors.  In effect, the Advanced LIGO recycling cavities
incorporate beam expanding telescopes that reduce the sensitivity to
thermal lenses in the ITMs.  In a similar change, the Fabry-Perot arm
cavities have a near concentric configuration motivated by the reduced
coating thermal noise from large spot sizes, discussed in
\S\ref{sec:therm}, and from the improved response to optical torques
for near concentric cavities, discussed in
Ref.~\citep{Sidles:2006p4148}.

The Advanced LIGO differential arm length will be detected in a
homodyne scheme known as DC readout. In DC readout, the GW signals are
measured directly as amplitude modulations on a static field at the AS
port.  The static field is created by a small offset in the
differential arm length; GWs make oscillations around the offset,
modulating the output.  However, many fields are present at the AS
port that don't carry GW information such as auxiliary control fields
and scattered non-resonant light.  Advanced LIGO incorporates an
Output Mode Cleaner (OMC) at the AS port to select only those fields
containing GW signal.  The OMC is a $\sim1~$m long optical cavity
which filters the interferometer output before detection, transmitting
only light from the arm cavity.

Finally, the Initial LIGO laser will be upgraded from a 10~W Master
Oscillator/Power Amplifier (MOPA) to a 180~W MOPA for Advanced
LIGO.~\citep{Willke:2008p10173} The input optics are upgraded to match
the laser: high power versions of electro-optic phase modulators,
photodetectors, and Faraday isolators replace conventional
components. With these changes, the maximum Advanced LIGO arm power
approaches 800~kW, improving the shot noise limited sensitivity by a
factor of $\approx 6$ with respect to Initial LIGO.

\section{Advanced LIGO Noise Contributions}
\label{sec:noiselimits}

\begin{figure}[tb]
  \centering
  \includegraphics[height=4in]{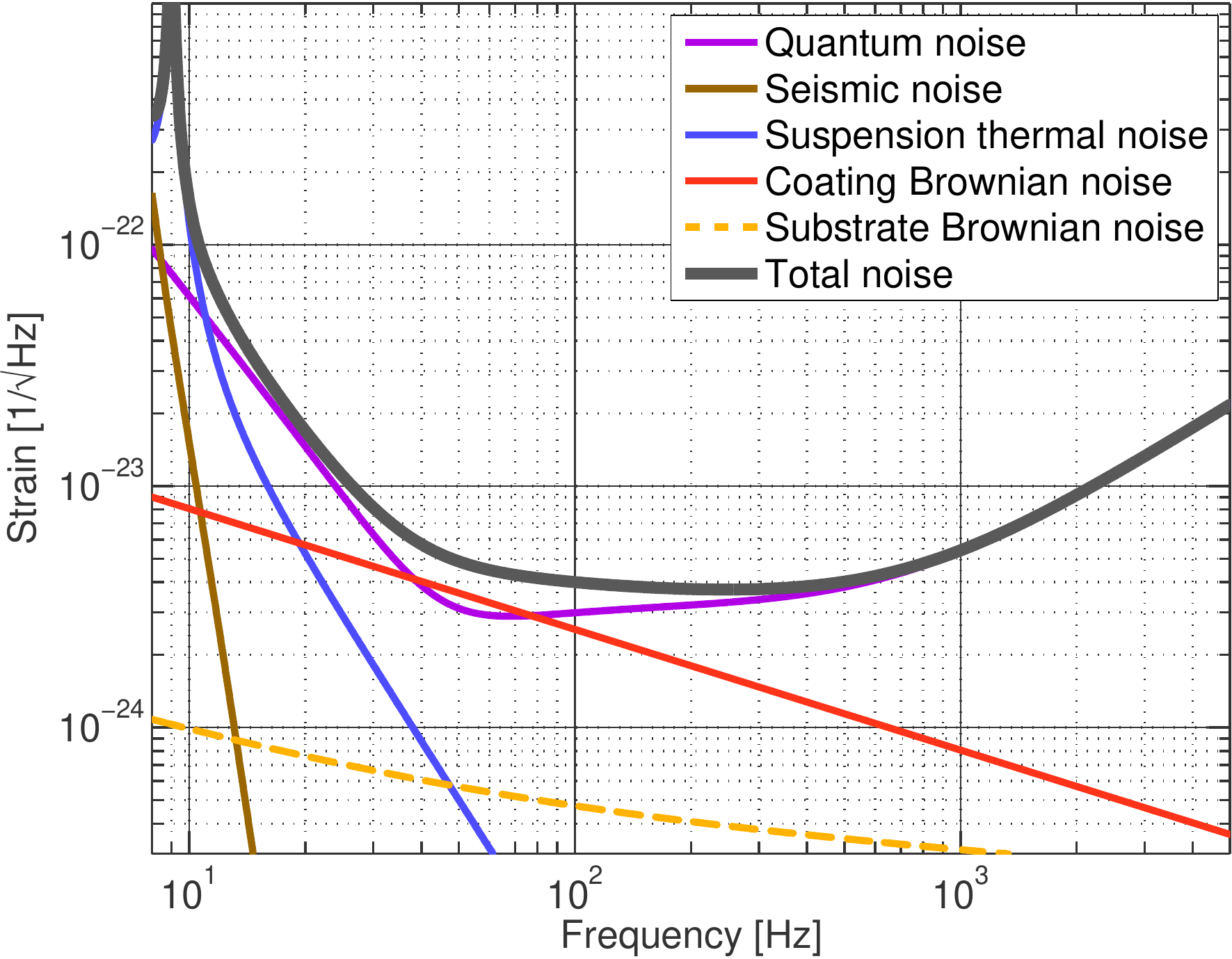}
  \caption{The modeled noise budget for an Advanced LIGO
    interferometer with $\phi_{SRM}= 0$ and 125~W input
    power.  The total noise (grey) is the incoherent sum of each of
    the listed noise terms,  described in detail below.}
  \label{fig:noise_budget}
\end{figure}

The Advanced LIGO sensitivity limits are estimated from calculations of
technical and fundamental noises; many of these have been studied with
Initial LIGO and other dedicated experiments.\footnote{See
  Ref.~\citep{Abbott:2009p9436} and references therein.}  Below 10~Hz,
the sensitivity is limited by the seismic motion of the earth, at
intermediate frequencies thermal noise dominates, and at the highest
frequencies photon shot noise limits the sensitivity. The
interferometer noise contributions, modeled with the GWINC-v2 software
package and plotted in Fig.~\ref{fig:noise_budget}, are described in
the following sections.

\subsection{Acoustic and Seismic Isolation}
\label{sec:isolation}

The Advanced LIGO detector requires a residual RMS differential arm
motion of $\delta x \leq 10^{-15}$~m to maintain the arm power buildup
and to minimize the coupling of laser noise into the GW signal.  In
the GW detection band, the required test mass displacement is $\delta
x \leq 10^{-19}~m/\sqrt{Hz}$ at 10 Hz and $\delta x \leq 2 \times
10^{-20}~m/\sqrt{Hz}$ at 100 Hz.  Here, $\delta x$ refers only to the
differential arm motion; the common arm motion and the motion of the
other length degrees of freedom may be somewhat larger.  To meet these
requirements, the interferometer is be isolated from environmental
influences such as acoustic noise, gas produced phase noise, and
seismic noise.  Thus, the interferometer optics are enclosed in an
ultra-high vacuum system.  The facility specifications have been
determined by the standard quantum sensitivity limit for a future
interferometer with 1 ton test masses.  The $\simeq 10^{-9}$~torr
vacuum reduces the gas produced phase noise to an equivalent strain of
$h \approx 10^{-25}$, well below the anticipated Advanced LIGO
sensitivity. In addition, the interferometer detection beam paths,
including the photodetectors, are enclosed within the vacuum on the
seismic isolation platforms to eliminate acoustic coupling and reduce
the motion of light scatterers.

\begin{figure}[tb]
  \centering
  \includegraphics[height=3in]{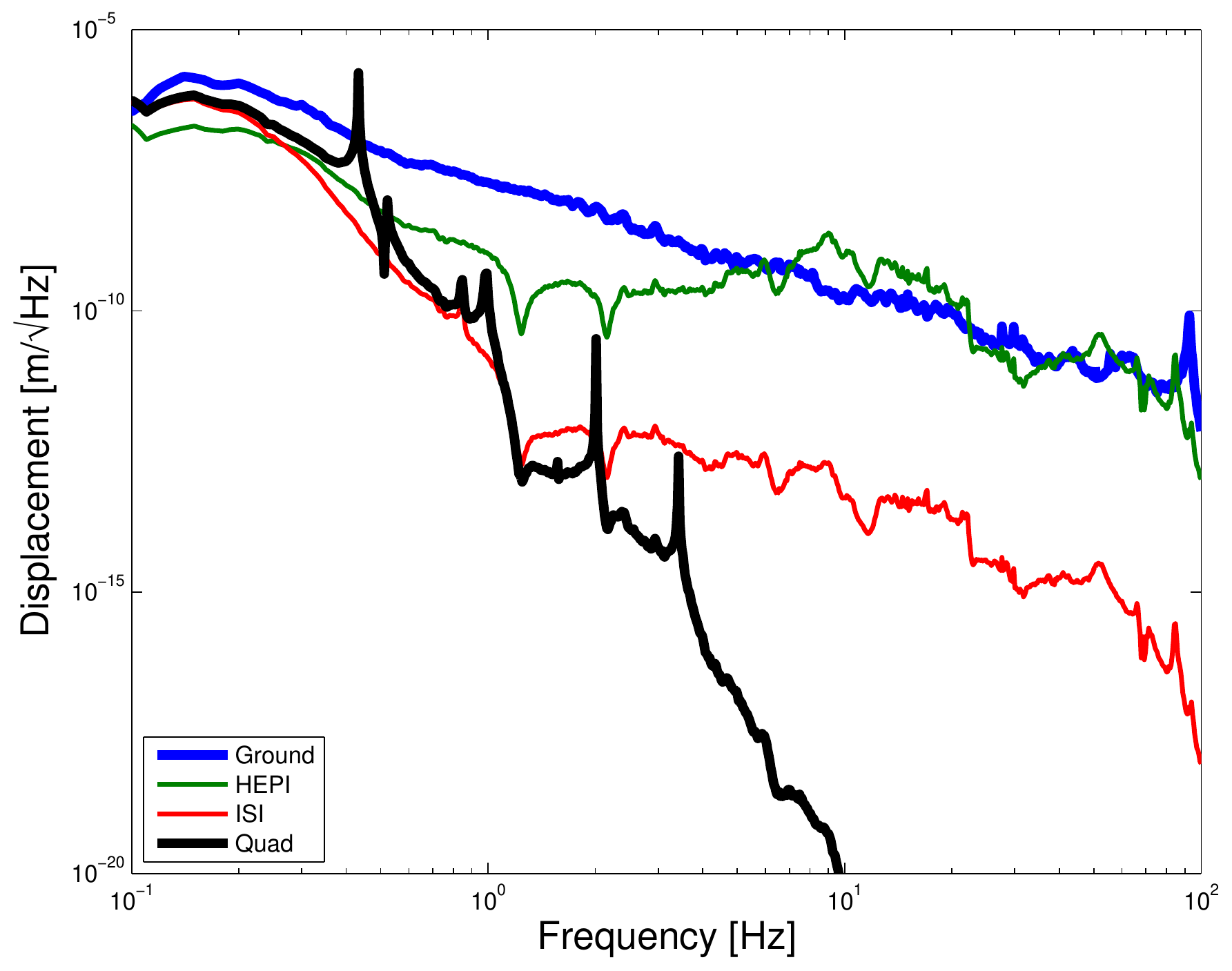}
  \includegraphics[height=3in]{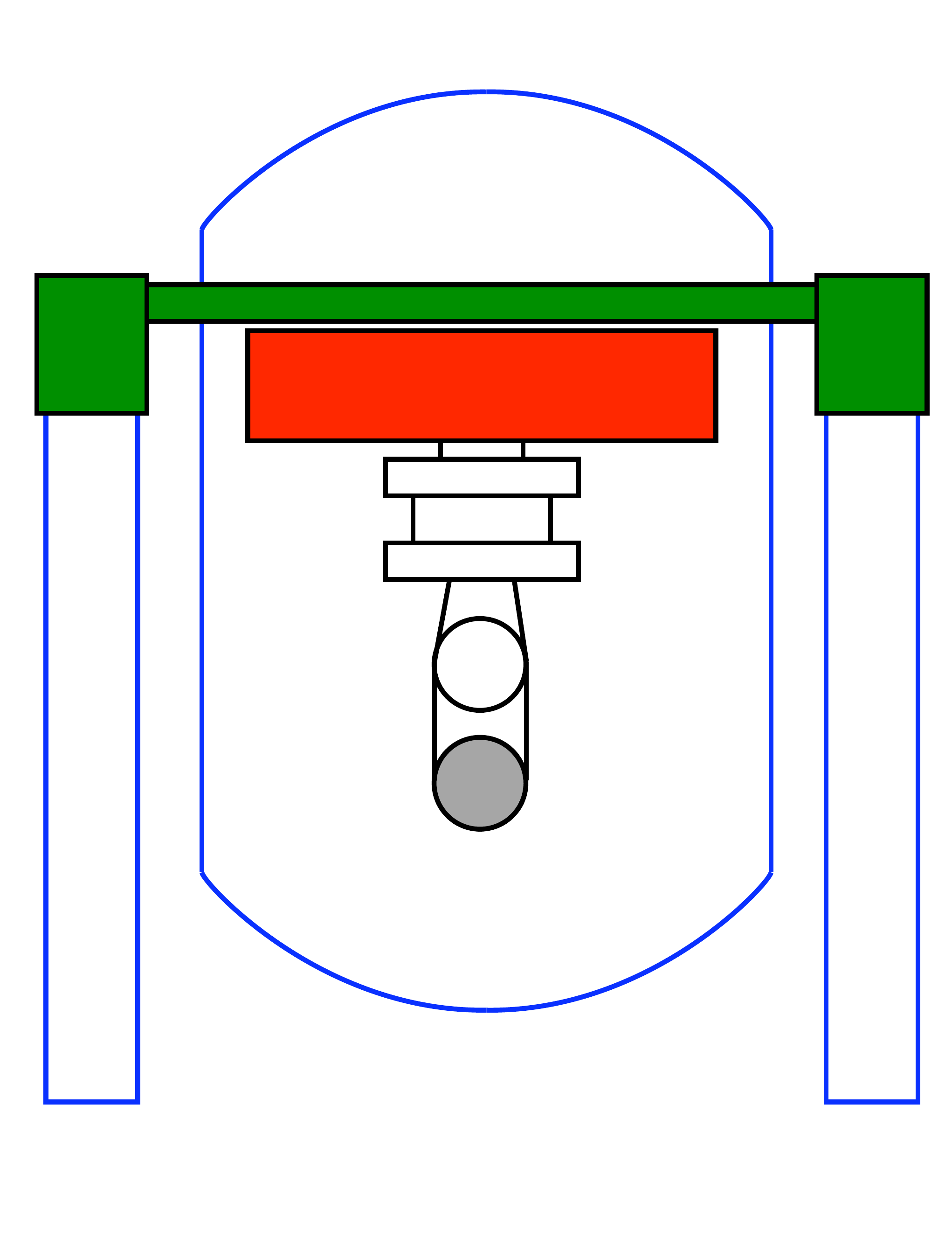}
  \caption{The Advanced LIGO seismic isolation consists of several
    stacked stages of active isolation shown schematically on the
    right, with colors matching the curve and the test mass shaded
    gray.  The reference ground motion (blue) is the average ground
    displacement at the Livingston observatory. The HEPI curve (green)
    is the motion atop the hydraulic pre-isolator located outside the
    vacuum system.  The anticipated ISI payload motion (red) is
    calculated from models of the two stage active isolator
    performance with the HEPI motion as the input spectrum.  The
    motion of the test mass at the end of the four-stage suspension is
    calculated from a model and shown in the Quad curve (black).}
  \label{fig:seismic}
\end{figure}

Isolating the interferometer optics from ground motion is a task
divided into several stages, with each stage providing isolation for
the following stage.  The effect of each stage of isolation can be
seen in Fig.~\ref{fig:seismic}.  At the lowest frequencies, an active,
6 degree of freedom, hydraulic, external pre-isolator (HEPI) reduces
the motion between 0.1 and 5~Hz by a factor of $\simeq
10$.~\citep{Fritschel:2004p9420} A two stage, active, internal seismic
isolation (ISI) system enclosed within the vacuum reduces ground
motion by a further factor of $\simeq 300$ at 1~Hz and $\simeq 3000$
at 10~Hz.~\citep{Abbott:2002p4431} Both active isolation systems consist
of a spring mounted, actuated platform outfitted with a suite of
motion sensors.  The sensors measure the platform motion and feedback
to the actuators, thereby stabilizing the platform to a level  limited by the
sensor noise floors and mechanical cross coupling.

The interferometer optics are suspended from the ISI platforms using a
coupled pendulum system based on the GEO-600 three-stage
suspensions.~\citep{Plissi:2000p9494} The beam splitter, power
recycling cavity and signal recycling cavity optics are hung from
three-stage suspensions consisting of two metal masses linked with
steel wire, followed by the optic itself.  Vertical isolation is
provided by cantilevered blade springs mounted to the metal masses.
Above the pendulum resonances, the triple suspension provides
isolation proportional to $1/f^6$.  The ITMs and ETMs are suspended
from four-stages consisting of two metal masses with vertical springs,
a fused silica intermediate mass, and the test mass. Welded fused
silica fibers join the test mass to the fused silica mass
above.~\citep{Robertson:2002p9490} The additional isolation stage is
necessary to meet the displacement noise goals at 10~Hz, and the fused
silica fibers reduce the thermal noise as discussed below.
Altogether, the Advanced LIGO isolation systems reduce the
seismic-induced test mass motion by 10 orders of magnitude to $\delta
x \simeq 10^{-20}~m/\sqrt{Hz}$ at 10 Hz, opening the frequency band
from 10 to 40~Hz for gravitational wave searches.
 
\subsection{Thermal Noise Sources}
\label{sec:therm}

Between 10 and 200 Hz, thermal noise sources limit the interferometer
sensitivity.  An unavoidable consequence of energy dissipation, thermal
noise is modeled by applying the fluctuation-dissipation theorem to
all aspects of the system which influence the motion or measurement of
the arm cavity test masses.   The fluctuation dissipation theorem is
closely related to Brownian motion, hence thermal noise associated
with mechanical motion is often called Brownian noise. Three sources
of dissipation dominate the Advanced LIGO thermal noise: mechanical
loss in each test mass leads to fluctuations in the mirrored surface
of the test mass; dissipation within the suspension fibers generates a
fluctuating force on the test masses; and losses within the mirror's
dielectric coating generate a fluctuating phase shift of the reflected
light.
 
The test mass mechanical loss determines the level of substrate
Brownian noise that causes fluctuations of the surface with respect to
the center of mass.  Early Advanced LIGO designs considered sapphire
test masses for their superior mechanical and thermal
properties. ~\citep{Rowan:2000p10031} Since then, fused silica test
masses have demonstrated sufficiently low loss and have been adopted
as the substrate material.  The Advanced LIGO test masses are 40~kg,
high purity, low-inclusion fused silica cylinders 34~cm in diameter.
To maintain the substrates' excellent mechanical properties, no lossy
materials contact the test mass (eg.  magnets).  The masses are
suspended via fused silica mounting blocks hydroxy-catalysis bonded to
each side.~\citep{Rowan:1998p10037} Fused silica fibers are welded to
the blocks and connect to the upper fused silica mass in a similar
fashion. The test masses are actuated using a non-contact, low-force
and low-noise electrostatic drive. As a result of these measures, the
test masses have very low loss (mechanical quality factors
exceeding $10^7$), and correspondingly low fluctuations, contributing
to the strain noise at $h \approx 3 \times 10^{-24} f^{-1/2}\;
Hz^{-1/2}$.

Loss in the mechanical structure supporting the test masses generates
fluctuating forces at the masses' suspension
points.~\citep{Gonzalez:2000p3745} The extremely low mechanical loss
needed to limit the fluctuations motivates the fused silica fiber
stage of the four-stage suspension.  Since the loss is dominated by
the bending regions near the fiber ends, the cylindrical fibers are
laser polished and drawn from fused silica rod with a carefully
controlled, variable diameter.  At the ends where the fiber is welded
to the test mass, the fibers have a large diameter to reduce flexing
of the (potentially higher loss) welded joints.  The fibers then taper
with an optimized profile so that the bending occurs predominantly in
a low loss region of the fiber.  Suspension thermal noise contributes
to the detector noise below 30~Hz, limiting the low-frequency
sensitivity to $h \approx  2\times 10^{-21}f^{-2}\; Hz^{-1/2}$.

Finally, thermal noise from the dielectric mirror coatings limits the
detector noise between 40 and 140~Hz, the most sensitive region.  The
coatings are alternating layers of SiO$_2$ and titanium-doped
TaO$_2$.\citep{Harry:2007p9489} Both thermo-optic and mechanical loss
contribute to the thermal noise.  The thermo-optic noises include
thermo-refractive fluctuations in the layers' index of refraction as
well as the thermo-elastic fluctuations that modify the layer
thickness and hence the magnitude and phase of the reflected field.
The coating mechanical loss dominates the thermal noise by an order of
magnitude, primarily in the thick ETM high reflector.  Because the
thin film coatings have a high mechanical loss relative to the
substrate, the coating Brownian noise exceeds that of the substrate by
nearly an order of magnitude.  The coating noise is inversely
proportional to the beam diameter, motivating large spot sizes on the
optics as described in \S\ref{sec:advanced-ligo-design}. Coating
thermal noise limits the detector sensitivity to $h\approx 2.5 \times
10^{-23} f^{-1/2}\; Hz^{-1/2}$. 

\subsection{Quantum Noise}
\label{sec:quantum}

\begin{figure}[tb]
  \centering
  \includegraphics[height=3.5in]{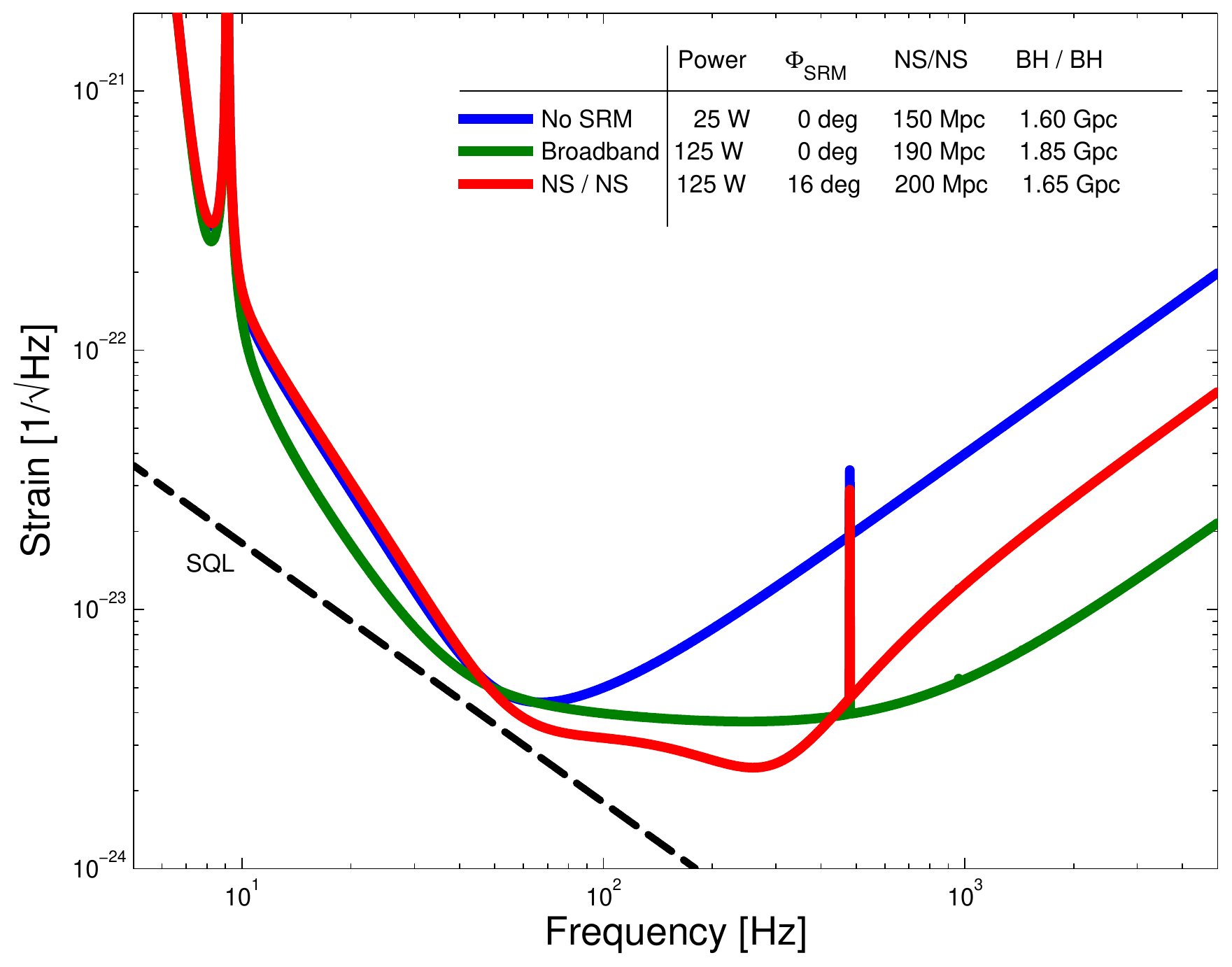}
  \caption{Amplitude spectral densities of the anticipated
    sensitivities of the Advanced LIGO interferometers as a function
    of the tuning of the signal recycling phase, $\phi_{SRM}$.  \textbf{No
      SRM} is a potential initial interferometer configuration with no
    signal recycling mirror with modest sensitivity; the
    \textbf{Broadband} configuration has good sensitivity at all
    frequencies; and \textbf{NS/NS} is optimized for the detection of
    two coalescing 1.4~M$_\odot$ neutron stars. }
\label{fig:noise_curves}
\end{figure}

Quantum mechanics limits the precision at which the test mass
positions can be determined.  At high frequencies, photon shot noise
limits the sensitivity to $h \propto \sqrt{ f / P}$, while at low
frequencies radiation pressure limits the sensitivity to $h \propto
\sqrt{P}/f^2$. The Advanced LIGO interferometer is a realization of a
Heisenberg microscope: the high laser power required to determine the
position of the test masses exerts a fluctuating radiation pressure
which perturbs the test mass positions.  In the absence of
position-momentum correlations, the Advanced LIGO strain sensitivity
is limited by the Standard Quantum Limit (SQL) $h_{SQL} = 1.8 \times
10^{-22}/f\; Hz^{-1/2}$.  Because the signal recycling cavity couples
the test mass position and momentum, sub-SQL sensitivity is possible
over a frequency bandwidth $\Delta f \sim f$ at the expense of the
sensitivity at other wavelengths.~\citep{Buonanno:2001p2} The primary
difference between the three curves shown in
Fig.~\ref{fig:noise_curves} is the degree of correlation between the
test mass position and momentum as determined by the SRM reflectivity
and the signal recycling cavity length tuning.\footnote{These noise
  curves can be found at
  https://dcc.ligo.org/cgi-bin/private/DocDB/ShowDocument?docid=2974
  and the documents therein.}  The noise curves include the thermal
and seismic noises calculated by GWINC-v2.

\section{Advanced LIGO Progress}
\label{sec:advanc-ligo-progr}
Some Advanced LIGO hardware has already been installed on the Initial LIGO
interferometers.  These components include: a 35~W laser
Master-Oscillator/Power-Amplifier; a high-power, in vacuum Faraday
Isolator; a single stage, in-vacuum seismic isolation system; and DC
readout using an in-vacuum Output Mode Cleaner.  With these systems,
LIGO has begun another science run, the sixth, with significantly
improved high frequency performance.
 
The Advanced LIGO project began in 2008, with plans for the first
in-vacuum hardware installation in early 2011.  To evaluate the
greatly increased chances for direct GW detection, we consider compact
binary coalescences for which the source rate can be estimated from
observed binary pulsar systems.  Once the instruments reach the
anticipated sensitivities, we can expect to detect between 1 and 1,000
compact binary coalescences per year. As installation and
commissioning progress, Advanced LIGO will transform the field from
searching for the first direct GW detection to exploring the rich
phenomena of GW astrophysics.

\section*{Acknowledgments}

We gratefully acknowledge the support of the United States National
Science Foundation for the construction and operation of the LIGO
Laboratory and the Science and Technology Facilities Council of the
United Kingdom, the Max-Planck-Society, and the State of
Niedersachsen/ Germany for support of the construction and operation
of the GEO600 detector.  We also gratefully acknowledge the support of
the research by these agencies and by the Australian Research Council,
the Council of Scientific and Industrial Research of India, the
Istituto Nazionale di Fisica Nucleare of Italy, the Spanish Ministerio
de Educacion y Ciencia, the Conselleria d'Economia Hisenda i Innovacio
of the Govern de les Illes Balears, the Scottish Funding Council, the
Scottish Universities Physics Alliance, The National Aeronautics and
Space Administration, the Carnegie Trust, the Leverhulme Trust, the
David and Lucile Packard Foundation, the Research Corporation and the
Alfred P Sloan Foundation.


\bibliographystyle{unsrt} 
\bibliography{blois_short} 
\end{document}